\documentclass[
aip,
amsmath,amssymb,
reprint,%
]{revtex4-2}

\usepackage[utf8]{inputenc}
\usepackage[english]{babel}
\usepackage{graphicx}% Include figure files
\usepackage{dcolumn}% Align table columns on decimal point
\usepackage{bm}% bold math
\usepackage{amsfonts,amssymb,amsmath}
\usepackage{siunitx}
\usepackage{hyperref}
\hypersetup{colorlinks=true,linktoc=all,linkcolor=blue,breaklinks=true,citecolor=blue,urlcolor=blue}

\newcommand{\SIi}{Supplementary Material}
\newcommand{\Qi}{$Q_\text{int}$}
\newcommand{\Qd}{$Q_\text{D}$}
\newcommand{\Qm}{$Q_\text{m}$}

\newcommand{\Qr}{$Q_\text{rad}$}
\newcommand{\Dq}{$D_\text{Q}$}
\newcommand{\dhh}{$d_\text{h}$}
\newcommand{\fd}{$f_\text{d}$}

\newcommand{\sr}{$\sigma_\text{released}$}

\newcommand{\sg}{$\sigma_\text{residual}$}

\newcommand{\fm}{$f_\text{m}$}

\newcommand{\Qf}{$Q_\text{m}\times f_\text{m}$}

\newcommand{\fref}[1]{Fig.~\ref{#1}}
\newcommand{\tref}[1]{Tab.~\ref{#1}}

\DeclareSIUnit\bar{bar}

\begin{document}
	
	\title{Membrane phononic crystals for high-\Qm{} mechanical defect modes in piezoelectric aluminum nitride}

	\author{Anastasiia Ciers}
	\email{anastasiia.ciers@chalmers.se}
	\author{Laurentius Radit Nindito}
	\author{Alexander Jung}
	\author{Hannes Pfeifer}
	\affiliation{Department of Microtechnology and Nanoscience (MC2),\\
		Chalmers University of Technology, SE-412 96 G\"oteborg, Sweden}
	\author{Armin Dadgar}
	\author{Andr\'e Strittmatter}
	\affiliation{Institute of Physics, Otto-von-Guericke-University Magdeburg, 39106 Magdeburg, Germany}%
	\author{Witlef Wieczorek}%
	\email{witlef.wieczorek@chalmers.se}
	\affiliation{Department of Microtechnology and Nanoscience (MC2),\\
		Chalmers University of Technology, SE-412 96 G\"oteborg, Sweden}

\begin{abstract}
	Nanomechanical resonators with exceptionally low dissipation are advancing mechanics-based sensors and quantum technologies. The key for these advances is the engineering of localized phononic modes that are well-isolated from the environment, i.e., that exhibit a high mechanical quality factor, \Qm{}. Membrane phononic crystals fabricated from strained thin films can realize high-\Qm{} single or multiple localized phononic defect modes at MHz frequencies. These defect modes can be efficiently interfaced with out-of-plane light or coupled to a microwave quantum circuit, enabling readout and control of their motion. When membrane phononic crystals are fabricated from a crystalline film, they could offer built-in functionality. 
	We demonstrate a membrane phononic crystal realized in a strained \SI{90}{\nano\meter}-thin film of aluminum nitride (AlN), which is a crystalline piezoelectric material. We engineer a high-\Qm{} localized phononic defect mode at \SI{1.8}{\mega\hertz} with a \Qf{}-product of \SI{1.5e13}{\hertz} at room temperature. In future devices, the built-in piezoelectricity of AlN can be utilized for direct coupling to qubits or in-situ tuning of mechanical mode frequencies, defect mode couplings, or acoustic bandgaps, which can be used as building blocks of tunable phononic circuits or low-noise sensors.
\end{abstract}

\maketitle

Periodic patterning of a material is a well-established method to engineer its effective properties.
Phononic crystals (PnC) are structures with periodic variations in the material's mechanical properties \cite{maldovan2013sound}  with the goal to engineer the localization \cite{eichenfield2009optomechanical,yu2014phononic,tsaturyan2017ultracoherent}  or propagation \cite{ren2022topological,ghasemi2018acoustic,xi2024soft} of mechanical excitations. 
PnC can be combined with engineering of the tensile strain distribution in a thin film. Tensile strain thereby allows storing mechanical energy in lossless tensile energy, a method known as dissipation dilution \cite{fedorov2019generalized,engelsen2024ultrahigh}. This combination results in soft-clamped, localized mechanical defect modes with ultra-high quality factors \cite{tsaturyan2017ultracoherent,reetz2019analysis,hoj2024ultracoherent}  in the kHz to MHz regime \cite{engelsen2024ultrahigh}, a frequency range relevant for sensing \cite{halg2021membrane} or microwave-to-optics transduction \cite{delaney2022superconducting}. 
Membrane-based, i.e., two-dimensional, PnCs thereby constitute a well-suited geometry, as they allow for efficient interfacing to out-of-plane light \cite{tsaturyan2017ultracoherent,brubaker2022optomechanical,clark2024optically}, coupling of MHz mechanical modes to GHz superconducting circuits \cite{brubaker2022optomechanical,seis2022ground,delaney2022superconducting}, or realization of coupled defect modes \cite{catalini2020soft}. Such membrane PnC structures have approached quality factors of $10^9$ at room temperature based on thin strained films of amorphous Si$_3$N$_4$, realized through subtractive \cite{tsaturyan2017ultracoherent,reetz2019analysis,saarinen2023laser} or additive density modulation \cite{hoj2024ultracoherent,huang2024room}. 
Although amorphous Si$_3$N$_4$ is an excellent material for high-\Qm{} nanomechanics, it lacks functionality. 

Crystalline films with large tensile strain can offer built-in functionality, such as piezoelectricity or superconductivity, and at the same time realize high-\Qm{} nanomechanical devices.
In particular, piezoelectricity can be used to actively modulate mechanical properties such as mode frequency, phononic bandgap, or coupling between localized phononic modes, which can be utilized in reconfigurable phononic circuits \cite{taylor2022reconfigurable} or for low-noise sensing \cite{rugar1991mechanical}.
Furthermore, piezoelectric materials allow realizing optoelectromechanical devices for microwave-to-optics conversion or for direct coupling to superconducting qubits, which has, for example, been demonstrated with GHz mechanics \cite{o2010quantum,vainsencher2016bi,jiang2020efficient,mirhosseini2020superconducting,jiang2023optically,weaver2024integrated}. 

High-\Qm{} nanomechanical resonators in the kHz to MHz frequency range have been demonstrated with the piezoelectric materials InGaP \cite{Cole2014,Buckle2018,manjeshwar2023high}, SiC \cite{sementilli2025low,hochreiter2025monolithic}, or AlN \cite{ciers2024nanomechanical,ciers2024thickness}, of which AlN offers the largest piezoelectric coupling coefficient. However, the geometry of these devices in these piezoelectric materials was restricted to one-dimensional structures, such as uniform \cite{Buckle2018,ciers2024thickness,sementilli2025low,hochreiter2025monolithic} or PnC beams \cite{ciers2024nanomechanical}, or two-dimensional structures that do not exploit a PnC such as membrane \cite{Cole2014}, trampoline \cite{manjeshwar2023high} or hierarchical resonators \cite{ciers2024nanomechanical}. The realization of a PnC membrane-based nanomechanical resonator - leveraging piezoelectric materials and combining PnC shielding with soft clamping - would offer a long-coherence time resonator at MHz frequency that can be applied in low-noise sensing \cite{halg2021membrane}, transduction \cite{delaney2022superconducting,najera2024high}, or phononic circuitry \cite{taylor2022reconfigurable,xi2024soft}.

In our work, we realize high-\Qm{} mechanical defect modes in a membrane PnC in tensile-strained \SI{90}{\nano\meter}-thin piezoelectric AlN. We adapt a hexagonal PnC, a so-called dandelion, previously demonstrated with Si$_3$N$_4$ \cite{saarinen2023laser}. 

\begin{figure}[t!hbp]
	\centering
	\includegraphics[width=0.5\textwidth]{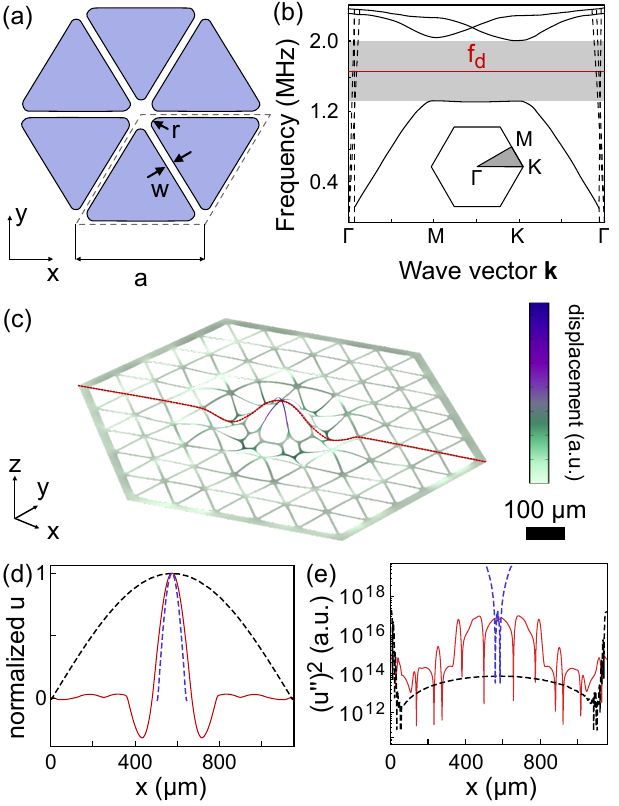}
	\caption{\textbf{Simulation of the unit cell and defect mode in a hexagonal PnC.} (a) PnC unit cell with lattice constant $a$ and triangular holes with fillet radius $r$ and tether width $w$. Dashed lines mark the primitive unit cell. (b) Phononic band structure of the unit cell showing a quasi-bandgap in the gray area. Solid lines are out-of-plane modes and dashed lines are in-plane modes. The red line indicates the defect-mode frequency \fd{} for \dhh{}$=\SI{45}{\micro\meter}$. The inset shows the irreducible Brillouin zone with the high-symmetry points $\Gamma$, M, and K.
		(c) Mechanical displacement of the defect mode. 
		(d) The displacement profile, $u$, and (e) squared curvature (linked to bending loss) for the defect mode along the red line marked in (c). Dashed lines in panels (d) and (e) show the prediction for a uniform beam of the same length (\SI{1.2}{\milli\meter}) as the PnC structure (black) and of a \SI{135}{\micro\meter}-long beam with a frequency of \SI{1.6}{\mega\hertz} that is similar to the defect-mode frequency (blue).
	}
	\label{fig:unit_cell}
\end{figure}

\fref{fig:unit_cell}(a) shows a unit cell of the hexagonal PnC lattice that we use to pattern the AlN film. A hexagonal PnC realizes a wider bandgap compared to a square-lattice PnC \cite{kushwaha1996giant,kuang2004effects} and, thus, isolates a PnC defect mode more effectively \cite{wang2017harnessing}. Furthermore, the hexagonal lattice has \SI{60}{\degree} in-plane rotational symmetry and, thus, follows the crystal structure and, therefore, stress distribution in AlN \cite{ciers2024thickness}. This is advantageous as the hexagonal lattice avoids an undesired directional variation of the PnC bandgap.

\fref{fig:unit_cell}(b) shows the band structure of the first irreducible Brillouin zone of the PnC along the high-symmetry points $\Gamma$, M, and K (see the inset in \fref{fig:unit_cell}(b)). The band structure is obtained from finite element method (FEM) eigenfrequency simulations (details in \SIi). For simplicity, we simulate AlN as an isotropic material with a Young's modulus of $\SI{313}{\giga\pascal}$ \cite{ciers2024thickness}. We obtain a 40\% bandgap for out-of-plane modes around \SI{1.66}{\mega\hertz} (\fref{fig:unit_cell}(b)) with the chosen pattern of rounded triangular holes defined by the lattice period $a = \SI{110}{\micro\meter}$, tether width $w = \SI{6.4}{\micro\meter}$, and fillet radius $r = \SI{5}{\micro\meter}$ (see \fref{fig:unit_cell}(a)).
Note that a pattern with circular holes would also result in a large bandgap \cite{tsaturyan2017ultracoherent}. However, such PnC design does not withstand the isotropic release process that we use for releasing the AlN layer (see \SIi).

\begin{figure}[t!hbp]
	\centering
	\includegraphics[width=0.5\textwidth]{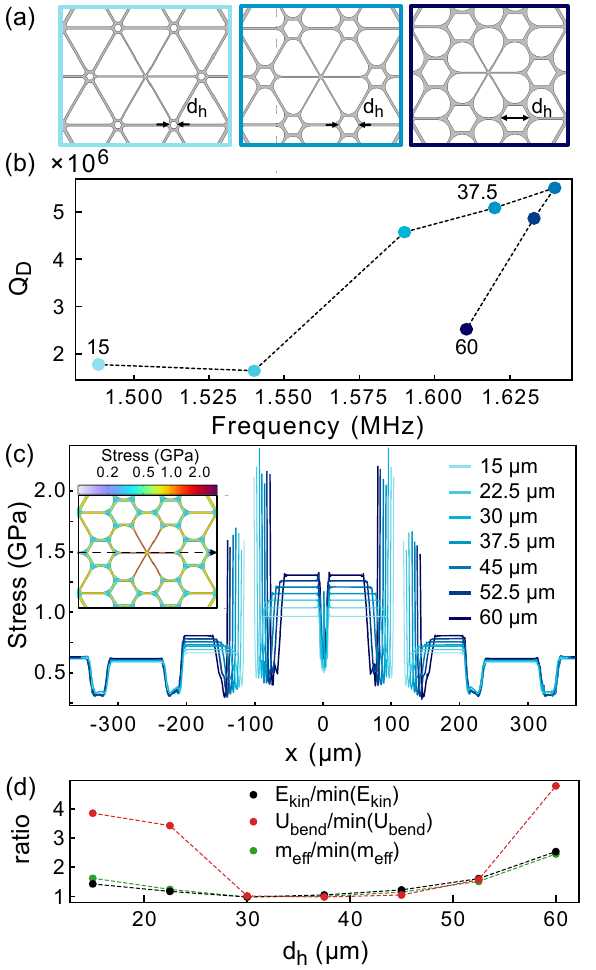}
	\caption{\textbf{FEM simulations for the PnC defect mode with different defect sites.} (a) Schematics of defect sites with \dhh{} = \SI{15}{\micro\meter}, \SI{37.5}{\micro\meter}, and \SI{60}{\micro\meter}. (b) \Qd{} vs.~frequency of the defect mode for different \dhh{}. (c) Stress along the tethers of the defect site and PnC (dashed line in inset) for different \dhh{}. The inset shows the stress distribution for the defect site with \dhh{} = \SI{60}{\micro\meter}. (d) The hole size \dhh{} influences the kinetic energy (black), bending energy (red), and the motional mass (green) of the localized defect mode. 
	}
	\label{fig:fem_dand_hole}
\end{figure}

We introduce a defect mode in the PnC by a variation of two unit cells from the center in all directions. The central defect pad has a diameter of \SI{8.7}{\micro\meter} and is supported by \SI{2.6}{\micro\meter}-wide tethers that taper to \SI{4.5}{\micro\meter} and then to $w = \SI{6.4}{\micro\meter}$ \cite{saarinen2023laser}. This results in a stress enhancement at the defect (see \fref{fig:fem_dand_hole}(c)), which is beneficial for increasing \Qm{}.
Similarly to Ref.~\cite{saarinen2023laser}, we patterned additional hexagonal holes of size \dhh{} in the pads directly adjacent to the central pad (see \fref{fig:unit_cell}(c)) leading to a gentler delocalization and, consequently, to a smaller bending of the mode, and also to the possibility to tune the frequency of the defect mode and optimize its quality factor. For example, for \dhh{} = \SI{45}{\micro\meter} the localized mode has a frequency of \SI{1.64}{\mega\hertz}, i.e., it is placed in the center of the PnC bandgap. \fref{fig:unit_cell}(d) shows the displacement profile of this defect mode, which is localized to the defect site, and strongly attenuated towards the clamping points due to the PnC. \fref{fig:unit_cell}(d) furthermore compares this displacement profile to the one of a \SI{1.2}{\milli\meter}-long beam, whose length is the same as the total PnC device, and to a \SI{135}{\micro\meter}-long beam, which has the same mechanical frequency as the defect mode of the PnC. We observe that the fundamental mode of the \SI{1.2}{\milli\meter}-long beam lacks the confinement that is achieved by the PnC membrane. 
\fref{fig:unit_cell}(e) shows the bending of the defect mode, which is much smaller at the clamping points compared to the uniform beams. Therefore, such a localized defect mode will exhibit a reduced bending loss.

To analyze the attainable \Qm{} of the defect mode, we consider that its total \Qm{} can be expressed as a sum of three dominant contributions (see \SIi{} for a discussion of other potential loss mechanisms):
\begin{equation}
	\frac{1}{Q_\text{m}} = \frac{1}{Q_\text{gas}} + \frac{1}{Q_\text{rad}} + \frac{1}{Q_\text{D}},
\end{equation}
where $Q_\text{gas}$ is the gas-limited quality factor (neglected in the following as we measure the devices in ultra-high vacuum), \Qr{} is the quality factor that is limited by radiation loss \cite{cole2011phonon,de2022mechanical} and \Qd{} is the diluted internal quality factor \cite{gonzfilez1994brownian}. 
The latter is given by
\begin{equation}
	Q_\text{D} = D_\text{Q} \times Q_\text{int}.
\end{equation}
The intrinsic quality factor for \SI{90}{\nano\meter}-thin AlN is \Qi{} = $7.4 \times 10^{3}$, which we determined from beams at MHz frequency in the same material, see \SIi{} and Ref.~ \cite{ciers2024thickness} for details. The dilution factor \Dq{} can be calculated from the ratio of the kinetic to the bending energy \cite{unterreithmeier2010damping,ciers2024nanomechanical}, $E_\text{kin}$ and $U_\text{bend}$, which we obtain from FEM simulations of the resonator geometry. 

We vary \dhh{} from 0 (no hole) to \SI{60}{\micro\meter} (\fref{fig:fem_dand_hole}(a)) to tune the defect mode's frequency and analyze the effect on \Qd{}. \fref{fig:fem_dand_hole}(b) shows that we obtain a non-monotonous behavior. The defect mode's frequency increases up to \dhh{}$=\SI{45}{\micro\meter}$ due to shortening of the tethers and an increased stress in the center region, see \fref{fig:fem_dand_hole}(c). Above \dhh{}$=\SI{45}{\micro\meter}$, the frequency decreases slightly, which we attribute to a small increase in the motional mass of the mode, see \fref{fig:fem_dand_hole}(d). To understand the behavior of \Qd{}, we notice that the defect-mode frequency is further away from the mid-bandgap frequency 
when \dhh{} is smaller than \SI{25}{\micro\meter}. This results in a weaker localization of the defect mode concomitant with an increase in bending energy (\fref{fig:fem_dand_hole}(d)). This results in a lower \Qd{} compared to the high \Qd{} achieved with \dhh{} between \SI{30}{\micro\meter} and \SI{55}{\micro\meter}. For \dhh{} of \SI{60}{\micro\meter} we observe a strong increase in the bending energy (\fref{fig:fem_dand_hole}(d)), which explains the observed reduction in \Qd{}.

\begin{figure}[t!hbp]
	\centering
	\includegraphics[width=0.5\textwidth]{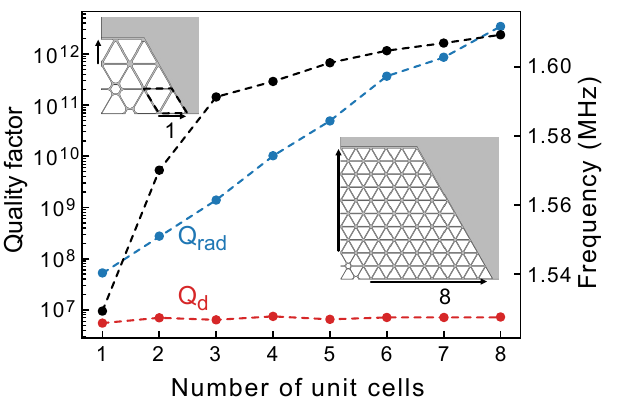}
	\caption{\textbf{Membrane PnC defect mode properties versus unit cell number.} Dissipation-diluted quality factor \Qd{} (red), radiation-limited quality factor \Qr{} (blue), and defect-mode frequency \fd{} (black) obtained from FEM simulations.}
	\label{fig:Q_n}
\end{figure}

To evaluate a potential limitation of \Qm{} by the radiation-limited quality factor \Qr{}, we analyze radiation loss. To this end, we vary the number of unit cells around the defect site from one to eight and evaluate loss of mechanical energy to the substrate via FEM simulations (see \SIi{}). 
\fref{fig:Q_n} shows that \Qr{} increases exponentially with the number of unit cells, which demonstrates efficient shielding, while \Qd{} remains approximately constant. Furthermore, we observe a slight increase in the defect-mode frequency, which we attribute to a stronger localization of the mode with an increase in the number of unit cells. 
Therefore, for our experiments, we choose to fabricate PnC devices with three rows of unit cells, which guarantees that these devices are not limited by \Qr{}. Furthermore, we select the PnC structures with \dhh{} = $(37.5, 45, 52.5)\SI{}{\micro\meter}$ as their defect mode frequency \fd{} is closest to the mid-bandgap frequency resulting in the highest values of \Qd{} (see \fref{fig:fem_dand_hole}(b)).

\begin{figure}[t!hbp]
	\centering
	\includegraphics[width=0.5\textwidth]{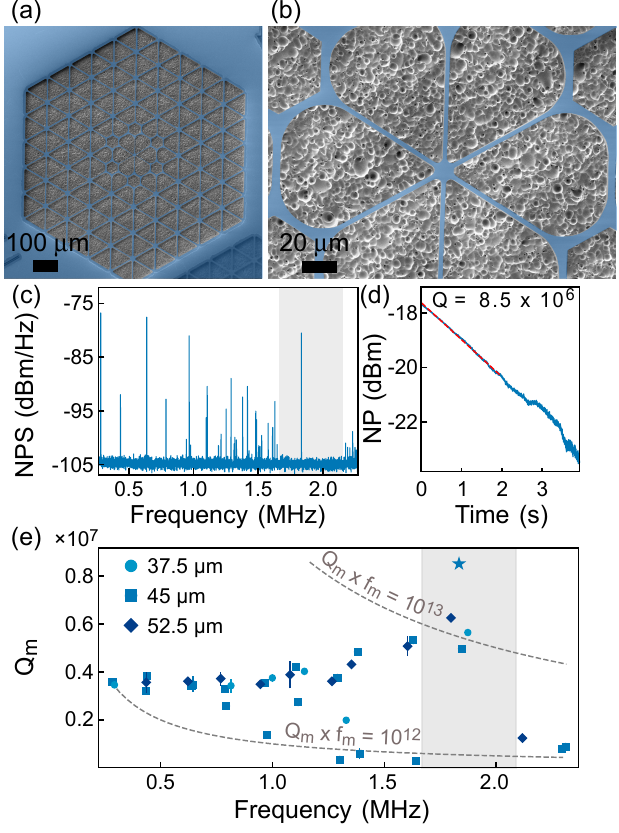}
	\caption{\textbf{Measurements of AlN dandelion membrane PnCs.} (a) False-colored SEM image of a released PnC device. (b) A close-up SEM image of the defect-site. (c) NPS of the membrane PnC with \dhh{} = \SI{45}{\micro\meter}. Gray area highlights the observed quasi-bandgap. (d) Ring-down measurement of the localized defect mode marked as a star in (e). (e) Measured frequencies and corresponding \Qm{} of non-localized and defect modes observed from four PnC devices with \dhh{} = (37.5, 45, 52.5)\SI{}{\micro\meter}. }
	\label{fig:Q_dand}
\end{figure}

We fabricate membrane PnC devices from an AlN film with a thickness of \SI{90}{\nano\meter} that is grown by metal-organic vapor-phase epitaxy (MOVPE) directly on a Si (111) substrate. This process yields a crystalline film comprising a \SI{70}{\nano\meter}-thick defect-rich region, topped by a \SI{20}{\nano\meter}-thick region exhibiting single-crystal-like structural quality (see \SIi{}). The film has a residual stress of \SI{790}{\mega\pascal} \cite{ciers2024thickness} and effective piezoelectric constant of \SI{1.1}{\pico\meter/\volt}, for details see \SIi{} and Ref.~\cite{ciers2024thickness}. We use electron-beam lithography and a subsequent XeF$_2$-based release process for fabrication of the membrane PnCs, for details on the fabrication and its potential scaling to wafer-size see \SIi{} and Ref.~\cite{ciers2024nanomechanical}. \fref{fig:Q_dand}(a) shows a scanning electron microscope (SEM) image of a fabricated \SI{90}{\nano\meter}-thin PnC device and \fref{fig:Q_dand}(b) shows the defect site in the PnC. We characterize the mechanical properties of the devices using an interferometric displacement measurement setup, where the samples are placed in ultra-high vacuum (\SI{4.6e-8}{\milli \bar}) at room temperature \cite{ciers2024nanomechanical,ciers2024thickness}. 
\fref{fig:Q_dand}(c) shows the noise power spectrum (NPS) of a PnC device with \dhh{} = \SI{45}{\micro\meter}. We observe a clear and well-isolated signal of the defect mode at a frequency of \SI{1.83}{\mega\hertz}. \fref{fig:Q_dand}(d) shows a ring-down measurement of this mode, from which we obtain a \Qm{} of $8.2 \times 10^6$, yielding a \Qf-product of $1.5 \times 10^{13}$\,\SI{}{\hertz}. The calculated force sensitivity for this defect mode using a simulated motional mass of \SI{0.4}{\nano\gram} at room temperature is $\SI{99}{\atto\newton/\sqrt{\hertz}}$, similar to Ref.~\cite{reetz2019analysis} (\SI{104}{\atto\newton/\sqrt{\hertz}}) and a factor of two larger than Ref.~\cite{tsaturyan2017ultracoherent} (\SI{55}{\atto\newton/\sqrt{\hertz}}), see \SIi{} for a more detailed comparison. 

\begin{table}[t!hbp]
	\centering
	\begin{tabular}{c|c|c|c|c}
		\hline
		\hline
		$d_\text{h}$ (\SI{}{\micro\meter}) & \fd{}$^\text{FEM}$ (\SI{}{\mega\hertz}) & \fd{}$^\text{meas}$ (\SI{}{\mega\hertz}) & \Qd{}$^\text{FEM}$ & \Qm{}$^\text{meas}$  \\
		\hline
		\hline
		37.5 & 1.62 &  1.87 & $5.1 \times 10^6$ & $5.6 \times 10^6$ \\ 
		45 & 1.64 &1.83 & $5.5 \times 10^6$ & $8.2 \times 10^6$ \\
		52.5 & 1.63 & 1.79 & $4.9 \times 10^6$ &  $6.2 \times 10^6$ \\
		\hline
		\hline
	\end{tabular}
	\caption{Simulated and measured values for the localized defect mode. Parameters: $a = \SI{110}{\micro\meter}$, $r = \SI{5}{\micro\meter}$.}
	\label{tab:fem_Qf}
\end{table}

\fref{fig:Q_dand}(e) summarizes the measured quality factors \Qm{} for the localized defect modes and non-localized modes of four fabricated devices. The largest quality factors and \Qf{}-products are observed for the localized defect modes, as expected. \tref{tab:fem_Qf} compares simulated and measured values for three defect-site geometries with different \dhh{}. We obtain a reasonable agreement between simulated \Qd{} and measured \Qm{}. This agreement supports that the fabricated membrane PnC structures in AlN are indeed limited by dissipation dilution. However, we find that the FEM simulations underestimate the defect mode frequency by 10\%.
This discrepancy may be due to a combination of effects that we could not account for in our FEM simulations, where we assume an isotropic and homogeneous material for simplicity. The thin film we use shows a small in-plane stress anisotropy of about 6\% \cite{ciers2024thickness} (see \SIi{}), which would result in a frequency variation of about 3\%. Furthermore, the varying material quality of our film results in an out-of-plane stain gradient across its thickness that impacts the frequency of the structure. Together, these effects could explain the observed discrepancy.

\begin{figure}[t!hbp]
	\centering
	\includegraphics[width=0.5\textwidth]{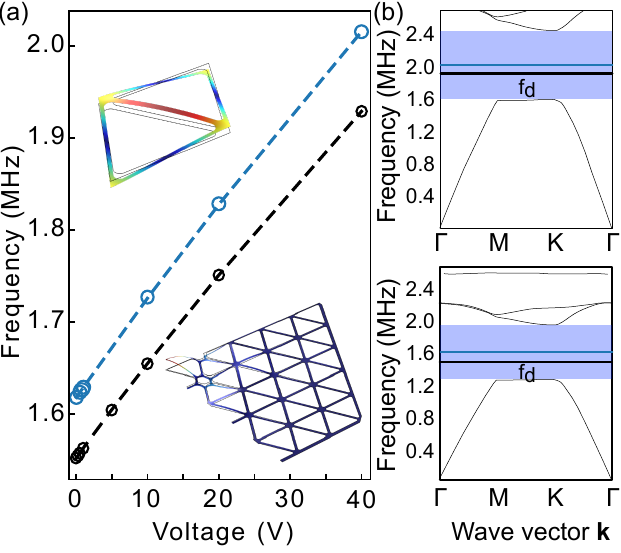}
	\caption{\textbf{Simulation of frequency-tuning capability of the PnC membrane.} Frequency vs.~applied voltage of the (a) center of the phononic bandgap (blue) and defect mode (black). (b) Band diagram highlighting the mid-bandgap frequency and the defect-mode frequency for a voltage of \SI{0}{\volt} (bottom) and \SI{40}{\volt} (top). }
	\label{fig:FEM_piezo}
\end{figure}

To conclude, we have realized dandelion \cite{saarinen2023laser} membrane phononic crystals in a crystalline \SI{90}{\nano\meter}-thin film of piezoelectric AlN. The membrane PnC supported a localized mechanical mode, where the best device reached a mechanical quality factor of $8.2 \times 10^6$ exploiting dissipation dilution via strain in the thin film and soft clamping via localization of the defect mode in the PnC. The \Qf-product of $1.5 \times 10^{13}$\,\SI{}{\hertz} supports a coherent mechanical oscillation at room temperature. Higher quality factors could be realized by improving the material quality or further engineering of the device geometry. By using thinner AlN films, higher dilution factors can be achieved provided that the strain and intrinsic quality factor of the film remain large. The membrane PnC device geometry can be further engineered by optimizing the PnC patterning \cite{zhang2017topological}, by exploiting strain engineering \cite{ghadimi2018elastic}, e.g. through further tapering of the tethers, or by exploring additive density modulation instead of etching holes \cite{hoj2024ultracoherent,huang2024room}.
Integration of doped AlGaN layers around the tensile-strained AlN film would allow harnessing the built-in piezoelectricity of the AlN film with an expected tunability of \SI{10}{\kilo\hertz/\volt} (see \fref{fig:FEM_piezo} and \SIi{}). Hybrid high-\Qm{} optoelectromechanical devices can then exploit this in-situ tunability for reconfigurable phononic circuits \cite{taylor2022reconfigurable}, topological phononics \cite{vasileiadis2021progress}, sensing \cite{halg2021membrane}, or microwave-to-optics conversion \cite{mirhosseini2020superconducting,jiang2020efficient,brubaker2022optomechanical,delaney2022superconducting,weaver2024integrated}.

The \SIi{} contains details on the material quality and fabrication, additional images of fabricated devices, parameters used in FEM simulations, a comparison to similar PnC devices, estimates of tunability and loss mechanisms.

We thank Joachim Ciers for valuable discussions and support with ellipsometry measurements.
This work was supported by the Knut and Alice Wallenberg (KAW) Foundation through a Wallenberg Academy Fellowship (W.W.), the KAW project no.~2022.0090, and the Wallenberg Center for Quantum Technology (WACQT, A.C.). H.P. acknowledges funding by the European Union under the project MSCA-PF-2022-OCOMM. MOVPE of AlN on Si was performed at Otto-von-Guericke-University Magdeburg. The mechanical resonators were fabricated in the Myfab Nanofabrication Laboratory at Chalmers. 
	
\section*{Author declarations}
\subsection*{Conflict of Interest}
The authors have no conflicts to disclose.

\section*{Data availability}
Data underlying the results presented in this paper are available in the open-access Zenodo database: \\ https://doi.org/10.5281/zenodo.14778128 \cite{zenododata}.

\bibliography{NanomechcrystalAlNhighQ.bib}

\clearpage
\begin{widetext}
\appendix

\section{Material characterization and fabrication}

\subsection{Crystal quality of AlN film }

The crystallinity and dislocation density of the AlN film were analyzed by high resolution X-ray diffraction (HRXRD) in Ref.~\cite{ciers2024thickness}. A narrower full width at half maximum (FWHM) of the diffraction peak corresponds to less disorder of the crystal plane. The twist distribution around the $c$-axis is evaluated via the (10$\Bar{1}$0) peak, \fref{fig:XRD}(a), while the $c$-axis disorder is evaluated using the $\omega$-scan of the AlN (0002) peak \fref{fig:XRD}(b). The AlN film with a thickness of \SI{90}{\nano\meter} exhibits a FWHM of the (0002) peak of \SI{0.7}{\degree} and a FWHM of the (10$\Bar{1}$0) peak of \SI{1.16}{\degree}. From \fref{fig:XRD}(c, d) we determine that $c = 4.9768$\,\AA{}, $a= 3.11663$\,\AA{}.
These values confirm high crystallinity of the AlN film.

\begin{figure}[h!tbp]
	\centering
	\includegraphics[width=\textwidth]{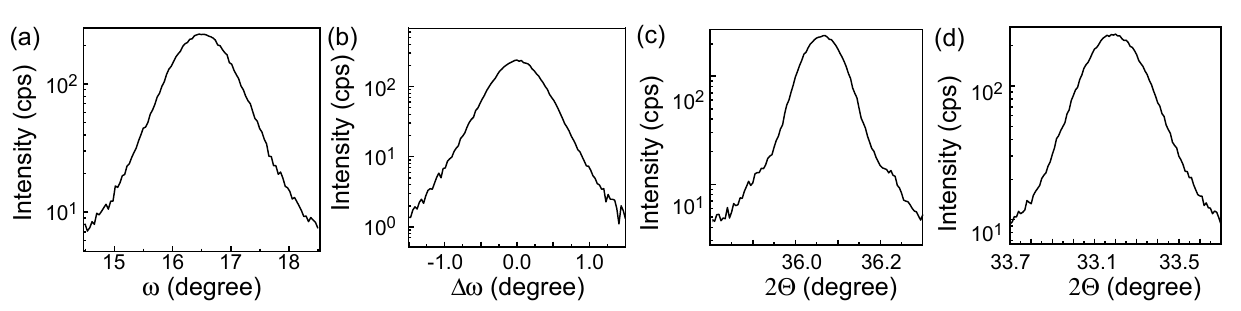}
	\caption{HRXRD measurements of \SI{90}{\nano\meter}-thick AlN film. (a) Grazing incidence in-plane diffraction (GIID) $\omega$-scan for (10$\Bar{1}$0), (b) $\omega$-scan at (0002), (c) $\Theta/\theta$-scan for (0002), (d) GIID 2$\Theta/\theta$-scan for (10$\Bar{1}$0).}
	\label{fig:XRD}
\end{figure}

\subsection{Piezoelectricity of AlN}

We use piezoresponse force microscopy (PFM) to determine the effective piezoelectric coefficient (see Refs.~\cite{ciers2024nanomechanical,ciers2024thickness} for details of the measurement procedure performed with a commercial Bruker Dimension ICON AFM). The deformation of the film induced by the electric field is detected by the PFM probe and the effective piezoelectric value of $d_{33}^\text{eff} = 1.1$\,pm/V is measured. This value is smaller than the bulk value due to the clamping constraint of the film through the substrate \cite{kim2005study}, the film's polycrystallinity in the first \SI{70}{\nano\meter}, and, potentially, the stress in the film \cite{berfield2007residual}.

\subsection{Released stress of AlN film}
\label{sec:stress}

The released stress of the \SI{90}{\nano\meter}-thick AlN film was extracted from beams of various length (\SI{75}{\micro\meter} to \SI{200}{\micro\meter}) and orientation on the sample (0 to \SI{180}{\degree}). The resulted stress anisotropy is plotted in \fref{fig:stress}.

\begin{figure}[h!tbp]
	\centering
	\includegraphics[width=\textwidth]{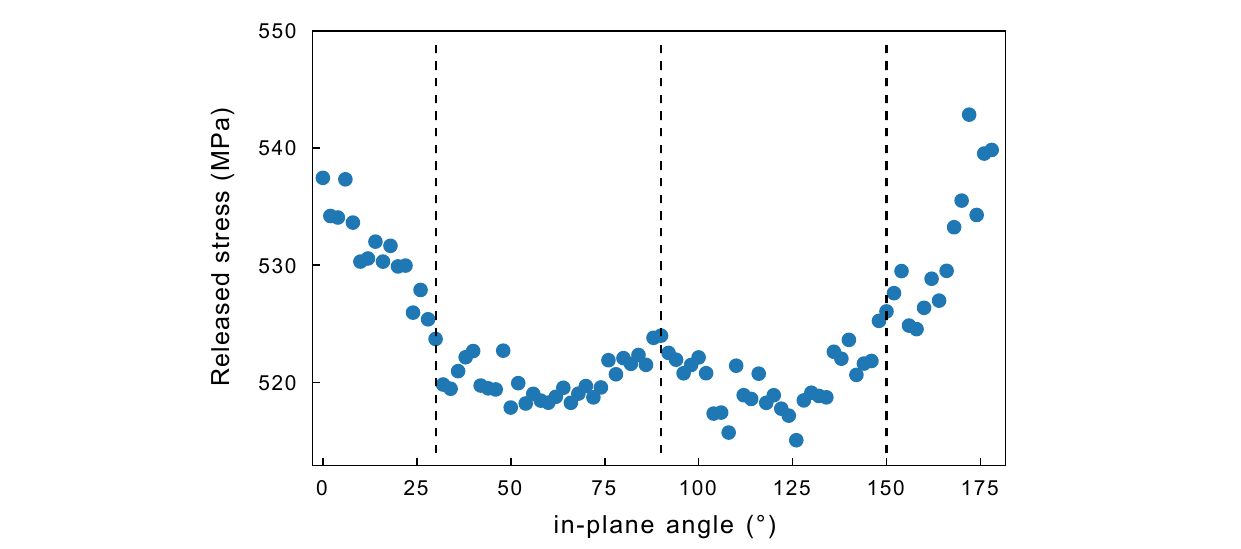}
	\caption{\textbf{In-plane released stress variation} of \SI{90}{\nano\meter}-thick AlN beams.}
	\label{fig:stress}
\end{figure}

\subsection{Determination of \Qi{}}

The intrinsic mechanical quality factor of the \SI{90}{\nano\meter}-thin AlN released film can be determined from measurements of the quality factor of uniform beams, see, e.g.~Ref.~\cite{unterreithmeier2010damping}. 
The quality factor of a strained, high-aspect-ratio doubly-clamped uniform beam is enhanced by the dilution factor, $D_Q$, over the intrinsic quality factor, \Qi{} as \cite{fedorov2019generalized}
\begin{equation}
	Q^{\mathrm{beam}}_\text{D} = D_Q^{\mathrm{beam}} \ Q_\text{int},
	\label{eq:dissipation_dilution}
\end{equation}
where the dilution factor is $D_{Q}^{\mathrm{beam}}=\left[{(\pi\lambda)^2 + 2\lambda}\right]^{-1}$ and the stress parameter $\lambda = \frac{h}{L} (12\text{\sr$/E$})^{-1/2}$ \cite{schmid2011damping}. In Ref.~\cite{ciers2024thickness} we measured beams with a length between \SI{75}{\micro \meter} to \SI{200}{\micro \meter} and determined their stress \sr$=\SI{520}{\mega\pascal}$ (see Sec.~\ref{sec:stress}) and  \Qi{} = $7.4 \times 10^3$. We refer to Ref.~\cite{ciers2024thickness} for details of the measurements.

\subsection{Fabrication}

\fref{fig:opt_dand} shows an optical image of a successfully fabricated dandelion membrane PnC in \SI{90}{\nano\meter}-thin AlN.

\begin{figure}[h!tbp]
	\centering
	\includegraphics[width=\textwidth]{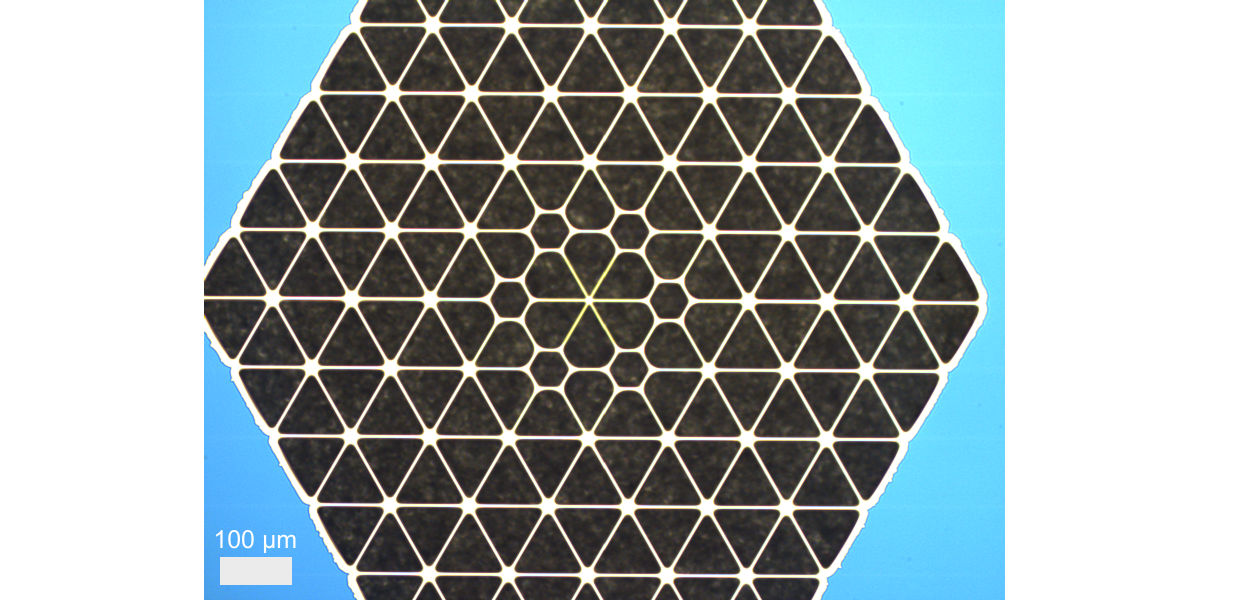}
	\caption{\textbf{Optical image of a dandelion membrane PnC with rounded triangular holes fabricated in \SI{90}{\nano\meter}-thick AlN.} \dhh{} = \SI{37.5}{\micro\meter}. }
	\label{fig:opt_dand}
\end{figure}

The release of the mechanical resonator is performed by dry etching in XeF$_2$, for details on the fabrication see Refs.~\cite{ciers2024nanomechanical,ciers2024thickness}. Such etching is reported to suffer from non-uniformity across the wafer, see, e.g., Ref.~\cite{chang1995gas}. As this process does not involve plasma, it heavily depends on the gas flow distribution, which requires a proper chamber design. However, there are techniques to improve etching homogeneity, such as to introduce pulse etching \cite{chang1995gas}, which one can consider if fabrication on the wafer-scale is desired.
However, it is worth mentioning that some resonator designs are sensitive to the undercut at the clamping points. This is, for example, the case for non-localized high-\Qm{} modes, like the fundamental mode of trampoline resonators \cite{manjeshwar2023high} or hierarchal structures \cite{bereyhi2022hierarchical}. Contrary to this, localizing the desired mode through a PnC shield avoids by design sensitivity to the undercut at the clamping of the PnC membrane. Therefore, the release with XeF$_2$ is perfectly suited for the latter type of mechanical resonator designs. Hence, it should be possible to fabricate large (mm- to cm-long) structures, potentially
even on the wafer-scale.

\section{Finite element method simulations}

For all FEM simulations we start by performing a stationary solution to evaluate the redistributed stress in the PnC. Then we simulate the eigenfrequencies of the mechanical modes. The parameters used in the FEM simulation are given in \tref{tab:fem_param}. 
For bandstructure simulations Bloch-Floquet periodicity is imposed on the boundaries along the in-plane directions, and the wave vector $k$ is swept across the first Brillouin zone.
\begin{table}[h!tbp]
	\centering
	\begin{tabular}{c|c|c|c}
		\hline
		\hline
		Symbol &  Description & Value & Unit\\ \hline
		$\rho$ & density & 3255 &$\SI{}{\kilo\gram/\meter}^3$\\
		$\nu$  & Poisson's ratio & 0.287& \\
		$h$  & layer thickness & 90 &\SI{}{\nano\meter}\\
		\sg{} & as-grown stress & 0.79 &\SI{}{\giga\pascal} \\
		$E$ & Young's modulus & 313 &\SI{}{\giga\pascal}\\
		\Qi{} & intrinsic quality factor &$7.4 \times 10^3$&  \\
		\hline
		\hline
	\end{tabular}
	\caption{FEM simulation parameters of the \SI{90}{\nano\meter}-thin AlN film \cite{ciers2024thickness}.}
	\label{tab:fem_param}
\end{table}

\subsection{Piezoelectric tuning and electrical contacts}
We estimate the expected tunability of our PnC-membrane based on FEM simulations. To this end, we couple electrostatics with structural mechanics in a multiphysics FEM simulation to evaluate piezoelectrically-induced frequency tuning of the PnC and the defect mode. In the FEM model, a voltage is applied to the top surface of the structure, while the bottom surface is grounded. \fref{fig:Piezo_FEM} shows the result for tuning of the bandgap center frequency and the bandgap width (\fref{fig:Piezo_FEM}(a)), and tuning of the defect mode frequency (\fref{fig:Piezo_FEM}(b)). The center bandgap frequency and defect mode shift with a rate of about \SI{10}{\kilo\hertz/\volt}. The width of the bandgap is increasing with applied voltage, with about \SI{4}{\kilo\hertz/\volt}, concomitant with the increase of the center bandgap frequency. As bandgap center frequency and defect mode frequency shift at the same rate, the defect mode remains within in the bandgap, see \fref{fig:Piezo_FEM}(c). To independently tune the PnC center frequency and the defect mode would, thus, necessitate to pattern the top conductive layer and electrodes. The exact design of the electrodes we leave to future work. 

\begin{figure}[h!tbp]
	\centering
	\includegraphics[width=\textwidth]{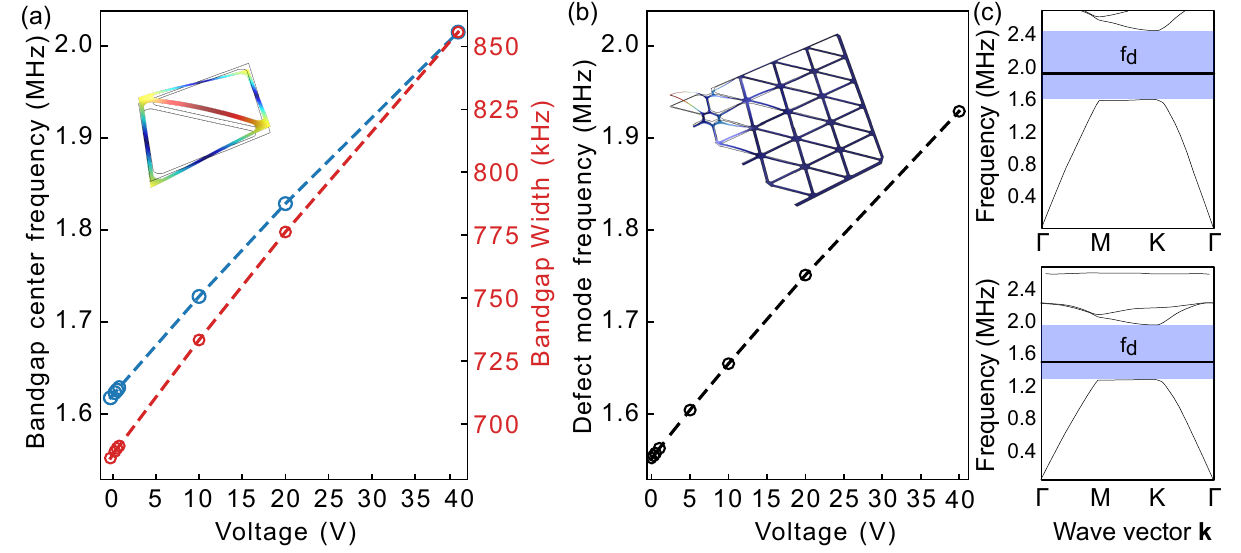}
	\caption{FEM simulation of the piezoelectric effect on the PnC-membrane structure. We show the frequency vs.~applied voltage characteristics of the (a) phononic bandgap and (b) defect mode. (c) Band diagram and the defect-mode frequency for a voltage of 0\,V (bottom) and 40\,V (top).}
	\label{fig:Piezo_FEM}
\end{figure}

To contact to the AlN layer, one can integrate doped AlGaN layers around the tensile-strained AlN film that would allow applying an electric field across the structure. To establish electrical contacts to the conductive AlGaN layers requires forming Ohmic contacts. These can rely on established work from the field of UV-LEDs based on AlGaN structures, see, e.g., Ref.~\cite{torres2024ultraviolet}, which utilized contacts to AlGaN layers for electrical driving of  LEDs. 

\section{Comparison: Force sensitivity and \Qf{} product}

We calculate the force sensitivity as $S_{F} = \sqrt{4 k_B T m_\text{eff} \Gamma_m}$ ($k_B$ is Boltzmann's constant, $\Gamma_m=\omega_m/Q_m$). 
The motional mass $m_\text{eff}$ is evaluated with respect to a point-like laser probe centered on the  maximum of the displacement field:
\begin{equation}
	m_\text{eff} = \frac{\rho \iint u^2 dx dy}{\max(u^2)}.
\end{equation}
For the distributed modes $m_\text{eff}$ is between 1 and \SI{10}{\nano\gram}, while the localized mode is below \SI{1}{\nano\gram}, e.g. $m_\text{eff} = $ \SI{0.43}{\nano\gram} for \dhh{} = \SI{45}{\micro\meter}. \tref{tab:fem_Qf_} shows $S_{F} $ for the defect modes of our devices. 

\begin{table}[h!tbp]
	\centering
	\begin{tabular}{c|c|c|c|c}
		\hline
		\hline
		$d_\text{h}$ (\SI{}{\micro\meter}) &  $m_\text{eff}$ (\SI{}{\nano\gram}) & \fd{}$^\text{meas}$ (\SI{}{\mega\hertz}) & \Qm{}$^\text{meas}$  & $S_{F}$ (\SI{}{\atto\newton/\sqrt{\hertz}}) \\
		\hline
		\hline
		37.5 &  0.37 &  1.87 &  $5.6 \times 10^6$ & 114 \\%111.2\\ %D = 686.69
		45 & 0.43 & 1.83 & $8.2 \times 10^6$ & 99.4 \\ %115 \\
		52.5 &  0.56 & 1.79 & $6.2 \times 10^6$& 129.7 \\%139\\
		\hline
		\hline
	\end{tabular}
	\caption{FEM values of $m_\text{eff}$ in dependence on \dhh{} ($a = \SI{110}{\micro\meter}$, $r = \SI{5}{\micro\meter}$). The FEM simulation values were obtained with the parameters from \tref{tab:fem_param}.}
	\label{tab:fem_Qf_}
\end{table}

\tref{tab:comp} compares our device to key parameters from similar devices. These devices operate in the tens of kHz to MHz range, which is relevant for force sensing. We compare  devices that have geometries like trampolines, phononic-shielded membranes, or hierarchically-clamped trampoline-like structures that operate at room temperature. These types of geometries are similar in the sense that they allow one to functionalize the central pad, such as with a photonic crystal or to realize coupled mechanical modes for force sensing. Further, operation at room temperature facilitates avoiding the use of cryostats.

\begin{table}[h!tbp]
	%    \centering
	\begin{tabular}{ccccccccc}
		\hline
		\hline
		Material & Thickness & $Q_m$ & $f_m$ &  $Q_m \times f_m$ & $m_\text{eff}$ & $S_F$ & Type & Ref\\
		& (nm) &  & (MHz)  & (Hz) & (ng) & (aN/$\sqrt{\text{Hz}}$) & \\  
		\hline
		\multicolumn{8}{l}{Amorphous materials}\\\hline
		Si$_3$N$_4$ & 20 & $2.3 \times 10^8$ & 0.1 &  $2.3 \times 10^{13}$ & 0.26 & 3.7 & hierarchal trampoline & \cite{bereyhi2022hierarchical}\\
		Si$_3$N$_4$ & 41 & $2.6 \times 10^7$ & 1.4 &$3.7 \times 10^{13}$ & 14 & 280 & PnC membrane & \cite{halg2021membrane}\\ 
		Si$_3$N$_4$ & 20 & $9.8 \times 10^7$ & 0.14 &$1.4 \times 10^{13}$ & 1 & 10 & trampoline & \cite{norte2016mechanical}\\ 
		Si$_3$N$_4$ & 15 & $1.4 \times 10^8$ & 1.29 &$1.8 \times 10^{14}$ & 0.2 & 13.9 & PnC membrane & \cite{saarinen2023laser}\\ 
		Si$_3$N$_4$ & 20 & $1.8 \times 10^8$ & 1.15 &$2.08 \times 10^{14}$ & 7.2 & 69.7 & PnC membrane & \cite{huang2024room}\\         
		\hline
		\multicolumn{8}{l}{Crystalline materials}\\\hline
		AlN & 90 & $8.5 \times 10^6$ & 1.8 &$1.5\times 10^{13}$ & 0.4 & 99 & PnC membrane  & this work\\
		AlN & 290 & $2.9 \times 10^7$ & 0.09 &$2.5\times 10^{12}$ & 2.67 & 29.2 & triangline  & \cite{ciers2024nanomechanical}\\
		SiC & 260 & $1.74 \times 10^6$ & 0.21 &$3.7 \times 10^{11}$ & 2.3 & 167.7 & trampoline  & \cite{romero2020engineering} \\
		InGaP & 73 & $1.8 \times 10^7$ & 0.04 & $7\times 10^{11}$ & 9.3 & 55 & trampoline  & \cite{manjeshwar2023high}\\
		\hline
		\hline
	\end{tabular}
	\caption{Comparison of the \Qf{}-product and force sensitivity for high-$Q_m$ mechanical resonators of trampoline-like or PnC-membrane type at room temperature.}
	\label{tab:comp}
\end{table}

In general, for force sensing one aims at a low-damped and low-mass mechanical mode as the force noise is proportional to $S_F$. Our result of about $99$\,aN/$\sqrt{\text{Hz}}$ lies within the range of about 5 to 300\,aN/$\sqrt{\text{Hz}}$ of similar resonator geometries realized in SiN or crystalline materials. The force sensitivity is not the only benchmark of a force sensor as other factors need to be considered as well, such as ease of use, the possibility to functionalize the resonator, or efficient read-out. The specific application will then determine the most fitting resonator geometry.

For operating in the quantum regime, the $Q_m \times f_m$-product is one of the figures of merit. This product should be larger than the thermal decoherence rate $k_B\cdot T/h$ ($h$ is Planck's constant), which is at room temperature $6\cdot 10^{12}$\,Hz, i.e., our device fulfills this condition. This condition is one of the requirements for ground-state cooling and can be interpreted as the number of coherent mechanical oscillations before thermal decoherence kicks-in.

\section{Mechanical loss}

\subsection{Extrinsic loss}

Let us discuss external loss mechanisms, which are radiation (or support) loss and squeeze-film damping.

\subsubsection{Radiation loss}

To evaluate the radiation loss $Q_\text{rad}$, we use the complex eigenfrequency obtained from FEM simulations \cite{de2022mechanical}:
\begin{equation}
	Q_\text{rad} = \frac{\text{Re}[f_\text{d}]}{2\, \text{Im}[f_\text{d}]}.
\end{equation}

The magnitude of radiation loss is dependent on both the impedance and mode matching
between the resonator and substrate, as well as the geometry of the resonator, and how the sample is mounted for the measurements.
To reduce the interference of reflected waves, the outer boundaries are defined as perfectly matched layers (PML). 
The evaluation of the radiation loss drastically depends on the setting of the PML.
The speed of the acoustic wave in AlN is $v = \SI{11}{\kilo\meter/\second}$ and in Si $v = \SI{8.4}{\kilo\meter/\second}$. Then, the wavelength for a frequency of \fm{} = \SI{1.5}{\mega\hertz} is given as $\lambda = v/$\fm{}, yielding $\lambda_\text{AlN} = \SI{7.33}{\milli\meter}$ and $\lambda_\text{Si} = \SI{5.6}{\milli\meter}$.
The PML should be somewhere between half and quarter of this wavelength (1.83 to \SI{3.67}{\milli\meter}) in order to attenuate the wave to a negligible amplitude before it reaches the outer boundary of the PML, while being short enough to keep computational costs reasonable.
We set the PML to \SI{3}{\milli\meter} and simulated a quarter of the actual structure (exploiting mirror symmetry). 
We introduce an additional transition layer (TL) of material between the clamping point of the PnC and the PML to avoid an abrupt change in material properties and therefore numerical artifacts. An empirical choice of the TL is about one eighth of the acoustic wavelength. \fref{fig:SI_PML}(a,b) shows the displacement obtained from such FEM simulations for the localized high-\Qr{} defect mode and a non-localized low-\Qr{} mode.

\begin{figure}[h!tbp]
	\centering
	\includegraphics[width=0.5\textwidth]{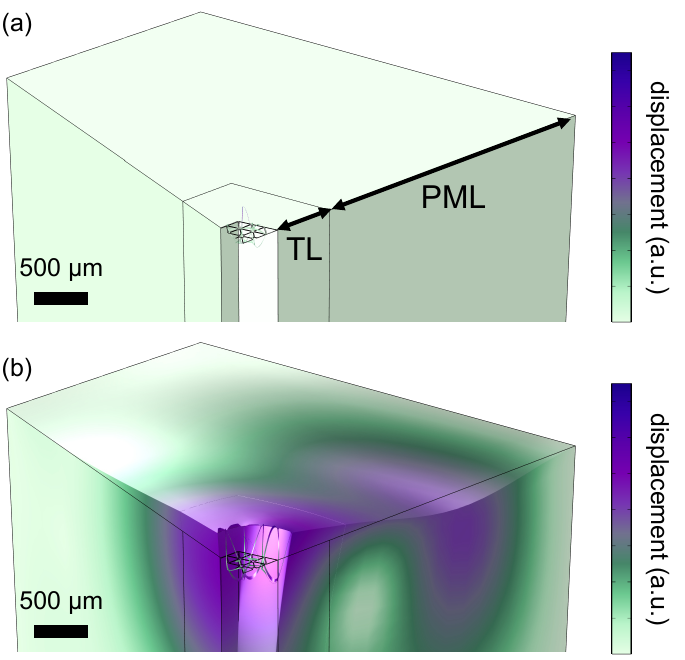}
	\caption{\textbf{FEM simulation of mode displacement.} (a) The defect mode and (b) low-\Qr{} mode with a perfectly matched layer (PML) and a transition layer (TL). }
	\label{fig:SI_PML}
\end{figure}

\subsubsection{Squeeze-film damping}

This loss mechanism comes into play when the resonator width is larger than the gap distance to the substrate. In our case the widest part of the PnC is about $w = \SI{10}{\micro\meter}$ while the distance to the substrate is at least $d_0 = \SI{30}{\micro\meter}$. To estimate squeeze-film damping, we use the equation that is valid in the ballistic regime from Ref.\cite{schmid2016fundamentals}:
\begin{equation}
	Q_\text{sf} = (2\pi)^{3/2}\rho h f_d \frac{d_0}{L_p} \sqrt{\frac{R_\text{gas}T}{M_m}}\frac{1}{p},
\end{equation}

whereby $\rho=\SI{3255}{\kilo\gram/\meter}^3$ is the density of AlN, $h=\SI{90}{\nano\meter}$ the thickness of the AlN layer,  $f_d = \SI{1.8}{\mega\hertz}$ the defect-mode frequency, $R_{gas}=\SI{8.314}{\joule}$\,K$^{-1}$mol$^{-1}$, $T = \SI{300}{\kelvin}$, and UHV environment with $p = 4.6\times 10^{-8}$\,mbar and $M_m = 2 \times 10^{-3}$\,kg/mol (residual gas at UHV is hydrogen, see Ref.~\cite{anufriev2006high}). 
As we have an extended two-dimensional PnC structure we estimate two cases: (i) assuming a beam with the parameters of the unit-cell tether only, i.e., with a peripheral length of $L_p=2\cdot (500+6.4)\,\mu$m, and (ii) assuming a square membrane without holes ($L_p=2\cdot (500+500)\,\mu$m).
We obtain a value of (i) $6.9 \times 10^{10}$ and (ii) $3.5 \times 10^{10}$, respectively. Therefore, $Q_\text{sf}$ of our localized defect mode should lie between these two values. As our measured $Q_m$ is close to $10^7$, we conclude that we are not limited by squeeze-film damping.

\subsection{Intrinsic loss}

We will now discuss loss mechanisms that are related to the material, which are bulk loss, such as Akhiezer and thermoelastic damping, and surface loss, which contribute to the intrinsic quality factor \Qi{}. Note that these loss mechanisms will be diluted by the dilution factor $D_Q$.

\subsubsection{Akhiezer damping} 

$Q_\textrm{Akh}$ is related to non-reversible heat flow under period deformation due to phonon scattering. Akhiezer damping starts to play a role at rather high frequency. The thermal relaxation time can be calculated as \cite{schmid2016fundamentals}
\begin{equation}
	\tau_\textrm{Akh}= \frac{3 \kappa}{c_p E} = 2.7 \text{\,ps},
\end{equation}
where $c_p$ is the specific heat capacity at constant pressure and $\kappa$ is the thermal conductivity (see \tref{tab:therm}). Therefore, $f_d \tau_\textrm{Akh} \ll 1$. In this limit one can evaluate $Q_\textrm{Akh}$ as \cite{ghaffari2013quantum}
\begin{equation}
	Q_\textrm{Akh}= \frac{\rho v^4}{2\pi \gamma_{avg}^2 \kappa T f_d} \approx 2\times10^7,
\end{equation}
temperature $T=300\,$K, $v$ is an average acoustic velocity \cite{dodd2001ultrasonic}, $\gamma_{avg} = 0.91$ is an average Gr\"uneisen parameter for AlN \cite{ghaffari2013quantum} (see \tref{tab:therm}). Hence, we do not expect yet to be limited by this mechanism.

\subsubsection{Thermoelastic damping} 

This loss mechanism is related to non-reversible heat flow under period deformation due to phonon transport. TED loss can be estimated based on the following formula assuming an unstrained beam (taken from Ref.~\cite{lifshitz2000thermoelastic}):
\begin{equation}
	\left(Q_\textrm{TED}(\xi)\right)^{-1} = \frac{\alpha^2 E T}{c_p}
	\left( \frac{6}{\xi^2} - \frac{6}{\xi^3}\frac{\sinh\xi + \sin\xi}{\cosh\xi + \cos\xi}\right),
\end{equation}
with the linear thermal expansion coefficient of AlN $\alpha = 3.5 \times 10^{-6}$ 1/K at a temperature of $T=300\,$K, $\xi(\omega_0) = h\sqrt{\frac{\omega_0}{2\chi}}$ with $\omega_0$ the isothermal value of the eigenfrequency $\omega$, $\chi = \kappa/\rho c_p = 0.88$ cm$^2$/s the thermal diffusivity. The parameters for AlN are in \tref{tab:therm} and \tref{tab:fem_param}.

However, we have a pre-strained film. Then, the value of quality factor limited by thermal relaxation is much higher. For instance, $Q_\textrm{TED,strain}$ can be calculated as \cite{kumar2010stress}:
\begin{equation}
	Q^{-1}_\textrm{TED,strain}  = \frac{1}{1+a\frac{F}{F_{cr}}}\left(Q_\textrm{TED}(\xi(\omega_0)\right)^{-1},
\end{equation}
where $F_{cr} = \pi EI/L^2$, $a = 0.97$ is a factor that depends
on the boundary condition, $I$ is the moment of inertia, and $\xi(\omega_0)$ gets modified by inserting $\omega_0=\frac{\pi}{L^2}\sqrt{\frac{EI}{\rho w h}}\sqrt{\pi^2+\frac{FL^2}{EI}}$, when we assume for simplicity that we have a beam. Using $L=135\,\mu$m, i.e., the length of a pre-strained beam that would yield the same resonance frequency as the defect mode, we obtain  $Q_\textrm{TED,strain}= 10^{11}$. Hence, TED does also not limit our observed quality factor.

\begin{table}[h!tbp]
	\centering
	\begin{tabular}{c|c|c}
		\hline\hline
		Parameter (units) & value & Ref. \\
		\hline
		Thermal expansion coefficient, $\alpha$ (\SI{}{1/\kelvin}) & $3.5\times10^{-6}$ & \cite{jackson1990temperature}\\
		Thermal conductivity, $\kappa$ (\SI{}{\joule/\kelvin}) & 208 & \cite{jackson1990temperature}\\
		Specific heat capacity, $c_p$ (\SI{}{\joule/\kilo\gram \kelvin}) & 725 & \cite{jackson1990temperature}\\
		Acoustic velocity, $v$ (\SI{}{\meter/\second}) & 11000 & \cite{dodd2001ultrasonic}\\
		Gr\"uneisen parameter, $\gamma_{avg}$  & 0.91 & \cite{ghaffari2013quantum} \\
		\hline\hline
	\end{tabular}
	\caption{Thermoacoustic and thermomechanical properties of AlN}
	\label{tab:therm}
\end{table}

\subsubsection{Surface loss} 

Surface loss will contribute to \Qi{} and is usually a limiting mechanism for thin films. In Ref.~\cite{ciers2024thickness} we extracted \Qi{} for different AlN film thicknesses. This \Qi{} has a contribution from the surface $Q_\text{surf}$ and the volume of the material $Q_\text{vol}$, i.e., $Q^{-1}_\text{int}=Q^{-1}_\text{surf}+Q^{-1}_\text{vol}$. Importantly, our AlN film consists of two different regions: defect-rich from the bottom surface of the film up to about 70\,nm and single-crystal-like from 70\,nm to 90\,nm, i.e., the top surface of the film. Each region has its own $Q_\text{vol}$ and $Q_\text{surf}$. In Ref.~\cite{ciers2024thickness} we estimated the average surface contribution as being linear and proportional to the thickness of the material with a slope of $4.8\cdot 10^{10}$1/m for thicknesses $h<200\,$nm. This value is slightly smaller than the one of SiN ($6\cdot 10^{10}$\,1/m \cite{villanueva2014evidence}). 
We believe that surface loss together with the polycrystallinity of the film gives the major contribution to \Qi{}. 

\section{PnC with circular holes}

Inspired by Refs.~\cite{tsaturyan2017ultracoherent,mason2019continuous}, we adapt a honeycomb PnC geometry for realizing a PnC in the AlN thin film.
\fref{fig:PnC_round}(a) shows a primitive unit cell of the hexagonal lattice with lattice constant $a$ and circular holes of radius $r$. 

\begin{figure*}[h!tbp]
	\centering
	\includegraphics[width=\textwidth]{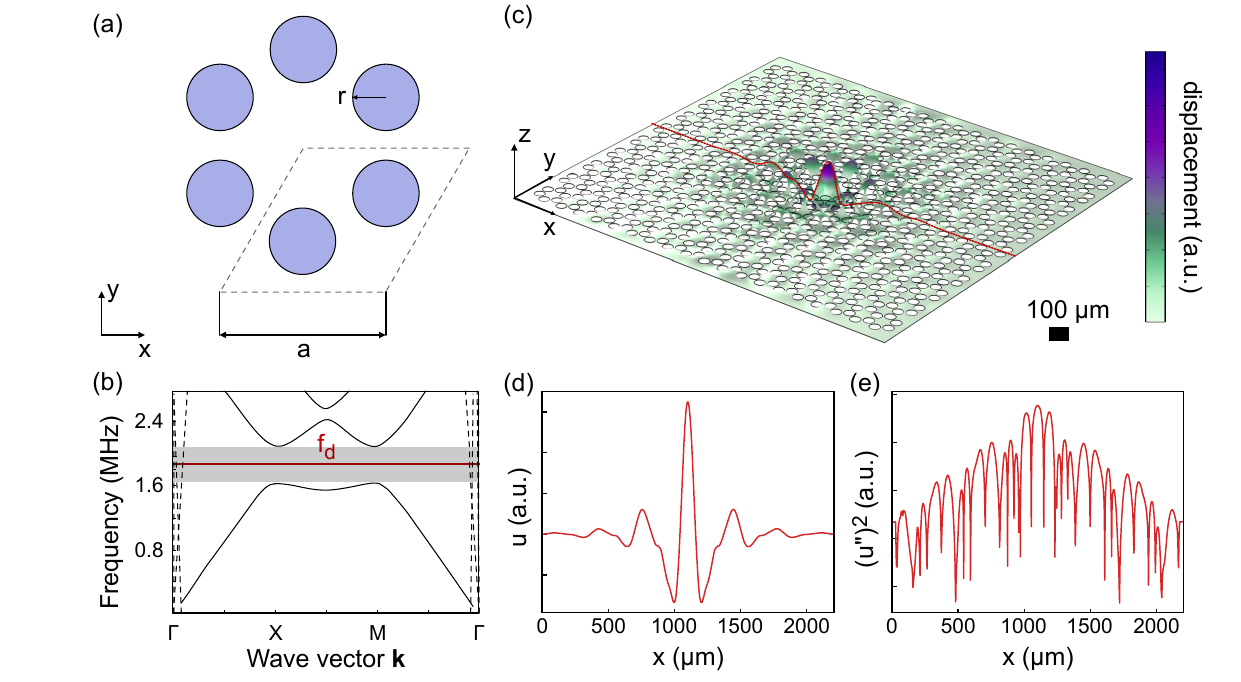}
	\caption{\textbf{FEM simulation of a honeycomb membrane PnC with circular holes.} (a) PnC unit cell with circular holes. (b) Phononic band structure of the unit cell. Solid lines are out-of-plane modes and dashed lines are in-plane modes. Grey area highlights the quasi-bandgap. 
		(c) Mechanical displacement of the defect mode in the PnC. 
		(d) The displacement and (e) curvature profiles along the red line marked in (c).
	}
	\label{fig:PnC_round}
\end{figure*}

\begin{table}[h!tbp]
	\centering
	\begin{tabular}{c|c|c|c|c}
		\hline
		\hline
		Hole type & $a$ (\SI{}{\micro\meter}) & $r$ (\SI{}{\micro\meter}) & \fd{}$^\text{FEM}$ (\SI{}{\mega\hertz}) &  \Qm{}$^\text{FEM}$ \\\hline
		Round & 110 & 0.245$a$ &  1.87 & $7.1 \times 10^6$ \\
		\hline
		\hline
	\end{tabular}
	\caption{Honeycomb PnC membrane simulation parameters and results.}
	\label{tab:fem_Qf_round}
\end{table}
For the parameters taken from \tref{tab:fem_Qf_round}, we obtain a bandgap for out-of-plane modes of around \SI{1.9}{\mega\hertz}. Similar to Refs.~\cite{tsaturyan2017ultracoherent,mason2019continuous} we remove and displace holes in the center of the PnC to break the symmetry of the lattice and form a defect in the PnC (\fref{fig:PnC_round}(c)), to which the mechanical mode of interest is confined, \fref{fig:PnC_round}(d). 
By adjusting the shape of the defect, we place the mechanical mode at a frequency of \SI{1.87}{\mega\hertz} and obtain a simulated \Qm{} = $7.1 \times 10^6$. 
We used eight unit cells in each direction from the center defect for this PnC geometry.

\begin{figure*}[ht!]
	\centering
	\includegraphics[width=\textwidth]{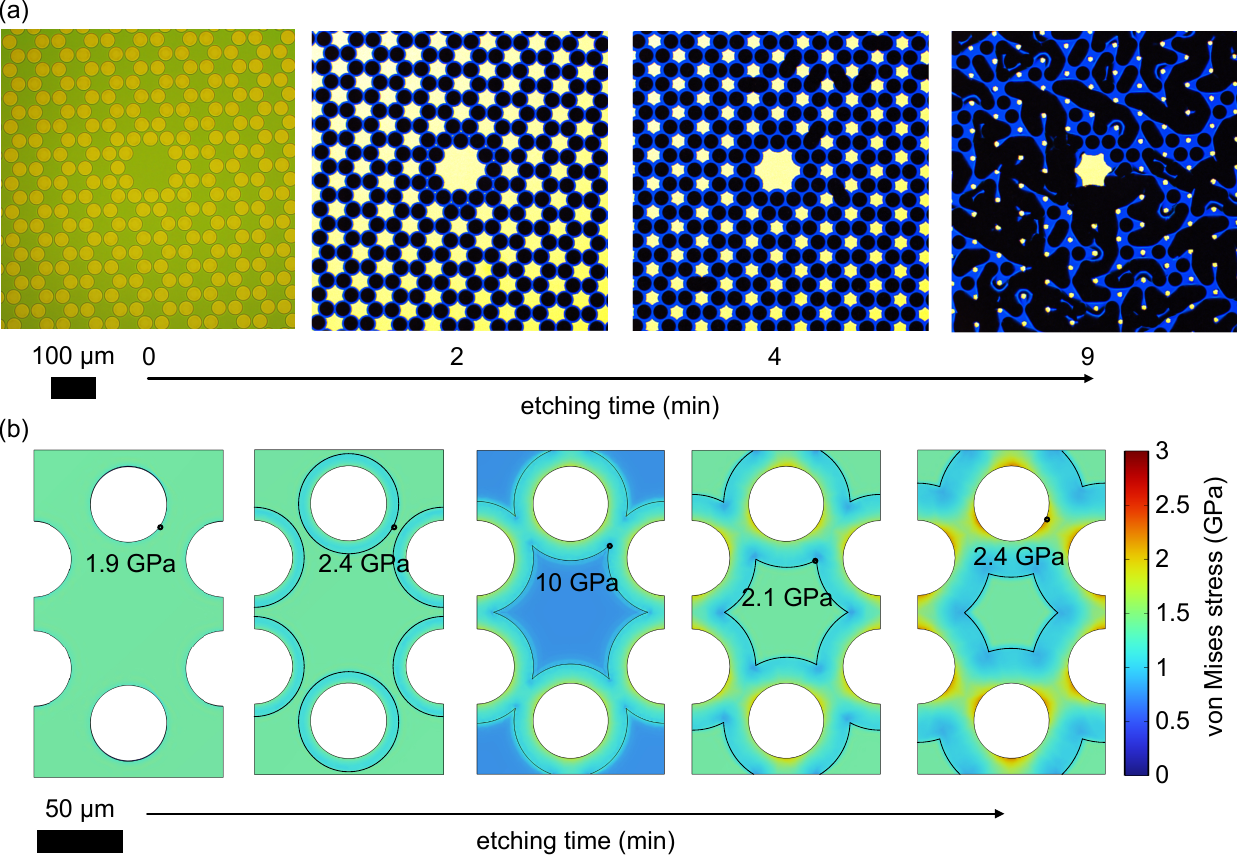}
	\caption{\textbf{Release process of the PnC with circular holes.} (a) Optical images of the honeycomb membrane PnC at different etching times. (b) FEM simulations indicating the maximum stress of the structure during the release process in the unit cell ($a = \SI{110}{\micro\meter}$).}
	\label{fig:h_etch}
\end{figure*}

When attempting fabrication of the hexagonal PnC with circular holes, it was not possible to release such structure successfully, see \fref{fig:h_etch}(a). 
To understand the unsuccessful release, we performed FEM simulations of the structure at different times of the fabrication, i.e., at different underetch lengths. \fref{fig:h_etch}(b) shows the stress in the unit cell at different etching times. We observe that at the moment when the tethers are suspended, the formed undercut has a sharp feature that causes the stress in the film to go above \SI{9}{\giga\pascal}. However, the exact value of the stress at the sharp undercut depends on the FEM meshing, which is used in the static solution simulation. Depending on this meshing, the stress can vary between 9 and \SI{20}{\giga\pascal}. We assume that this high value of stress is beyond the yield stress of the AlN film and, thus, leads to cracking and a failed release process.

\end{widetext}
	
\end{document}